\documentclass[preprint,12pt, a4paper]{elsarticle}

\usepackage{amssymb}
\usepackage{lineno}

\usepackage{float}
\restylefloat{table}

\usepackage{booktabs}% More professional look of tables.
\usepackage{multirow}% multirow on tables
\usepackage{hyphenat} % to better handle overfull hbox warnings
\usepackage{makecell} % for line break inside table cells
\usepackage{float} % for forcing figure in a place
\usepackage{tikz} % for flux charts
\usetikzlibrary{arrows.meta}
\tikzset{%
  >={Latex[width=2mm,length=2mm]},
            base/.style = {rectangle, rounded corners, draw=black,
                           minimum width=4cm, minimum height=1cm,
                           text centered, font=\sffamily},
  activityStarts/.style = {base, fill=blue!30},
       startstop/.style = {base, fill=red!30},
    activityRuns/.style = {base, fill=green!30, font=\ttfamily},
         process/.style = {base, minimum width=2.5cm, fill=orange!15,
                           font=\ttfamily},
           dashedrect/.style = {draw, dashed, fill=blue!5, thick, inner sep=3pt, minimum width=8em, font=\ttfamily},
}

\newcolumntype{L}[1]{>{\raggedright\arraybackslash}p{#1}}
 
\newcolumntype{C}[1]{>{\centering\arraybackslash}p{#1}}
 
\newcolumntype{R}[1]{>{\raggedleft\arraybackslash}p{#1}}

\usepackage[edges]{forest}

\definecolor{foldercolor}{RGB}{124,166,198}

\tikzset{pics/folder/.style={code={%
    \node[inner sep=0pt, minimum size=#1](-foldericon){};
    \node[folder style, inner sep=0pt, minimum width=0.3*#1, minimum height=0.6*#1, above right, xshift=0.05*#1] at (-foldericon.west){};
    \node[folder style, inner sep=0pt, minimum size=#1] at (-foldericon.center){};}
    },
    pics/folder/.default={20pt},
    folder style/.style={draw=foldercolor!80!black,top color=foldercolor!40,bottom color=foldercolor}
}

\forestset{is file/.style={edge path'/.expanded={%
        ([xshift=\forestregister{folder indent}]!u.parent anchor) |- (.child anchor)},
        inner sep=1pt},
    this folder size/.style={edge path'/.expanded={%
        ([xshift=\forestregister{folder indent}]!u.parent anchor) |- (.child anchor) pic[solid]{folder=#1}}, inner xsep=0.6*#1},
    folder tree indent/.style={before computing xy={l=#1}},
    folder icons/.style={folder, this folder size=#1, folder tree indent=3*#1},
    folder icons/.default={12pt},
}

\usepackage{caption}

\usepackage[capitalise]{cleveref} % for elsevier Fig. refs style

\usepackage{verbatimbox}
\usepackage[numbers]{natbib}

\journal{SoftwareX}

\begin{document}

\begin{frontmatter}



\title{mRpostman: An IMAP Client for R}


\author{Allan V. C. Quadros}
\ead{quadros@k-state.edu}

\address{Department of Statistics\\
    Kansas State University\\
    Manhattan, KS 66506, United States}

\begin{abstract}
Internet Message Access Protocol (IMAP) clients are a common feature in several programming languages. Despite having some packages for electronic messages retrieval, the R language, until recently, lacked a broader solution, capable of coping with different IMAP servers and providing a wide spectrum of features. mRpostman covers most of the IMAP 4rev1 functionalities by implementing tools for message searching, selective fetching of message attributes, mailbox management, attachment extraction, and several other IMAP features that can be executed in virtually any mail provider. By doing so, it enables users to perform data analysis based on e-mail content. The goal of this article is to showcase the toolkit provided with the mRpostman package, to describe its key features and provide some application examples.

\end{abstract}

\begin{keyword}
IMAP \sep e-mail \sep R



\end{keyword}

\end{frontmatter}

\section{Motivation and significance}
\label{motivation}

The acknowledgement of the R programming language\citep{R} as having remarkable statistical capabilities is much due to the excellence brought by its statistical and data analysis packages. This reputation also stands on the capabilities of a myriad of utility packages, which extends the use of the language by facilitating the integration of the steps involved in data collection, analysis, and communication. With that in mind, and considering the amount of data transmitted daily through e-mail, mRpostman was conceived to fill the absence of an Internet Message Access Protocol (IMAP) client in the R statistical environment; therefore, providing an appropriate toolkit for electronic messages retrieval, and paving the way for e-mail data analysis in R.

The Comprehensive R Archive Network (CRAN) has at least seven packages for sending emails (Table \ref{tab:r-packages}). Whereas some of these packages aim to provide a plain Simple Mail Transport Protocol (SMTP) client for R (e.g. sendmailR and emayili), others focus on more sophisticated implementations, using Application Program Interfaces (API), or providing seamless integration between SMTP and other R features such as rmarkdown\citep{rmarkdown}. However, despite the surplus of available clients in R, the SMTP protocol is not suitable for receiving e-mails. It only allows clients to communicate with servers to deliver their messages. 

For the purpose of message retrieval, there are the Post Office Protocol 3 (POP3) and the Internet Message Access Protocol (IMAP). In comparison with IMAP, POP3 is a very limited protocol, working as a simple interface for clients to download e-mails from servers. IMAP, on the other hand, is a much more complex protocol, and can be considered as the evolution of POP3, with a very different and broader set of functionalities. In contrast to POP3, all the messages are kept on the IMAP server and not locally. This means that a user can access the same mail account using parallel connections from different clients\citep{BOI08}. Besides the mail folders structure and management, the capacity of issuing sophisticated search queries also contribute to the level of complexity of the IMAP protocol.

Amid CRAN packages for e-mail communication, only gmailr and edeR have IMAP capabilities (Table \ref{tab:r-packages}). However, those capabilities are restricted to Gmail accounts and few IMAP functionalities. Although gmailr supports both protocols, the package is more SMTP-focused, which explains its low number of IMAP features. Therefore, R was clearly lacking a broader IMAP client solution. It was in that mainstay that mRpostman was conceived.

\begin{table}[H]
    \centering
    \resizebox{\textwidth}{!}{%
    \begin{tabular}{L{2.5cm}C{2.2cm}C{1.5cm}C{1.5cm}C{1.5cm}C{1.7cm}C{1.7cm}C{1.7cm}}
    \toprule
    & & & \multicolumn{4}{c}{Features} \\ \cmidrule(lr){4-7}
    &protocol &mail providers &search queries &message fetch &attachment extraction &mailbox management &active development\\ \midrule
sendmailR\citep{sendmailR} &SMTP &- &- &- &- &- &-\\
mailR\citep{mailR} &SMTP &- &- &- &- &- &-\\
mail\citep{mail} &SMTP &- &- &- &- &- &-\\
blatr\citep{blatr} &SMTP &- &- &- &- &- &-\\
gmailr\citep{gmailr} &SMTP/IMAP &Gmail &no &limited &limited &no &yes\\
blastula\citep{blastula} &SMTP &- &- &- &- &- &-\\
emayili\citep{emayili} &SMTP &- &- &- &- &- &-\\
edeR\citep{edeR} &IMAP &Gmail &no &limited &no &no &no\\
\textbf{mRpostman} &IMAP &all &yes &yes &yes &yes &yes\\
    \end{tabular}}
    \caption{Comparison of the current available CRAN packages for e-mail communication. The following attributes are evaluated: protocol - the supported protocol (SMTP or IMAP); mail providers - if the IMAP protocol is supported, which mail providers are supported by the package; Features - which type of IMAP features are available in the package; active development - if the package is currently under active development. If the package does not provide IMAP support, the remaining fields do not apply.}
    \label{tab:r-packages}
\end{table}

In this article, we present a brief view of the main functionalities of the package and its applications.

\section{Software description}
\label{description}

mRpostman is conceived to be an easy-to-use session-based IMAP client for R. The package implements intuitive methods for executing the majority of the IMAP commands described in the Request for Comments 3501\footnote{The RFC 3501\citep[]{rfc3501} is a formal document from the Internet Engineering Task Force (IETF) specifying standards for the IMAP, Version 4rev1 (IMAP4rev1).}, such as mailbox management, and selectively search and fetch of message attributes. The package also implements complementary functions for decoding quoted-printable and base 64 content, following the MIME specification\footnote{The RFC 2047\citep{rfc2047} specifies rules for encoding and decoding non-ASCII characters in electronic messages.}. 

All these methods and functions play an important role in facilitating e-mail data analysis. We shall not overlook the amount of data analyses daily performed on e-mail content. The package has proved to be very useful as an additional feature in this workflow by, for instance, enabling the possibility of automating the attachments retrieval step. Also, by fetching other message contents, users are able to apply statistical techniques for analysing the frequency of e-mails with regard to some message aspect, running sentiment analysis on e-mail content, etc.

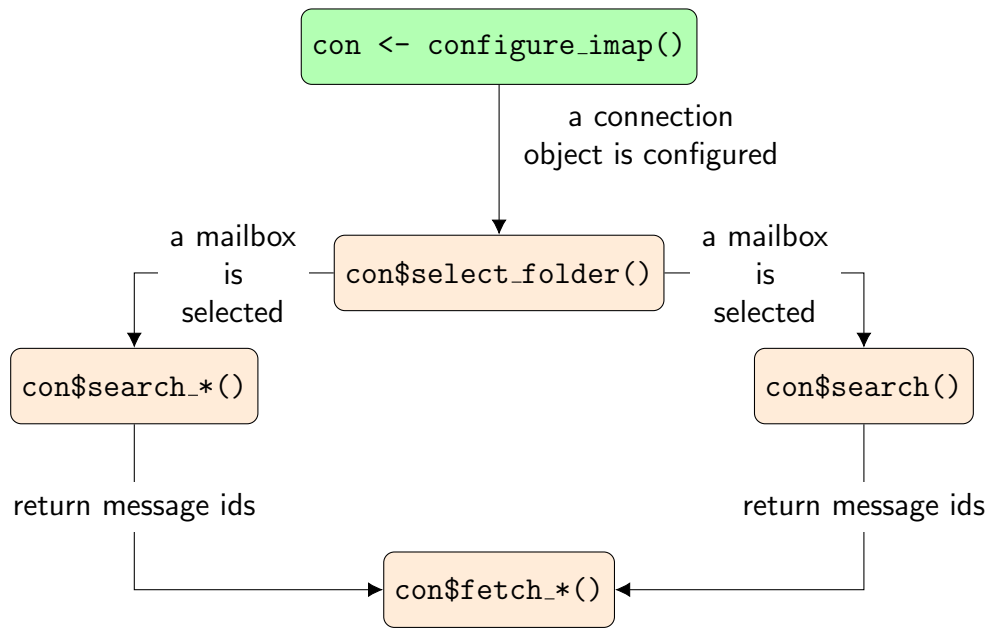
\begin{figure}[ht]
    \centering
    %\footnotesize
    \begin{tikzpicture}[node distance=1.5cm, every node/.style={fill=white, font=\sffamily}, align=center]
        %\label{fig:schema1}
  
        % Specification of nodes (position, etc.)
        \node (start)         [activityRuns]    {con <- configure\_imap()};
        %\node (capabilities)  [dashedrect, left of = start, yshift=-2.3cm, xshift=-2.7cm] {con\$list\_server\_capabilities()};
        \node (mbox)          [process, below of=start, yshift=-1.5cm]  {con\$select\_folder()};
        \node (fetch)         [process, below of = mbox, yshift=-2.7cm]   {con\$fetch\_*()};
        \node (regsearch)     [process, left of = fetch, xshift=-3.3cm, yshift=2.7cm]   {con\$search\_*()};
        \node (customsearch)  [process, right of = fetch, xshift=3.3cm, yshift=2.7cm]   {con\$search()};
        % Specification of lines between nodes specified above
        % with aditional nodes for description 
        \draw[->]             (start) -- (mbox) node[text width=3.5cm, yshift=1.8cm, xshift=2.0cm] {a connection object is configured};
        %\draw[dashed]         (capabilities) -- (0,-2.3) {};
        \draw[->]             (mbox) -| (regsearch) node[text width=1.7cm, yshift=1.5cm, xshift=1.3cm] {a mailbox is selected};
        \draw[->]             (mbox) -| (customsearch) node[text width=1.7cm, yshift=1.5cm, xshift=-1.3cm] {a mailbox is selected};
        \draw[->]             (regsearch) |- node[text width=4cm, yshift=1.1cm] {return message ids} (fetch) ;
        \draw[->]             (customsearch) |- node[text width=4cm, yshift=1.1cm] {return message ids}(fetch);
        
    \end{tikzpicture}
    \caption{Basic schema for fetching the full content of a message or its parts after a search query.}
    \label{fig:schema1}
\end{figure}

Since mRpostman works as a session-based IMAP client, one can think of the provided methods following a natural order in which the steps shall be organised in the event of an IMAP session (\cref{fig:schema1}). For instance, if the goal is to search messages within a specific period of time and/or containing a specific word, first we need to configure the connection to the IMAP server; then, choose a mail folder where the search is to be performed; and execute the single criteria (left) or the custom multi-criteria search (right). If the user intends to fetch the matched message(s) or its parts, additional fetch steps can be chained to the described schema. 

% Drawing part, node distance is 1.5 cm and every node
% is prefilled with white background

mRpostman is flexible in the sense that the aforementioned steps can be used either under the tidy framework, with pipes\citep{magrittr}, or via the conventional base R approach.

\section{Software architeture}
\label{architeture}

The software was designed following the object-oriented framework from the R6 package\citep{R6}. A class called \texttt{ImapCon} is implemented to retain and organize the necessary IMAP connection parameters. All the methods that derive from this class will serve one of the two following purposes: to issue a request toward the IMAP server (request methods) or re-configure an existing IMAP connection (reset methods).

In order to execute IMAP commands, this package makes extensive use of the curl\citep[]{curl20} R package\footnote{The curl package is a binding for the libcurl\citep[]{libcurl20} C library.}. All mRpostman's request methods are built on top of the so-called curl handles. Under the hood, a curl handle consists of a C pointer variable that gathers the necessary parameters to execute a request to the server. As a matter of fact, the handle itself does not issue any command, but is used as a parameter inside a curl's fetch function. This last object is the one that actually triggers the request to the server, ranging from mail folder selection to search queries, or message fetch requests. 

The object-oriented framework combined with the use of one curl handle per session enables mRpostman to elegantly run as a session based IMAP client, without demanding a connection reconfiguration between commands. For example, if a mail folder is selected on the current session, all requests using the same connection token will be performed on the selected folder, unless the user re-selects a different one.

\subsection{Software functionalities}
\label{functionalities}

%Provide at least one illustrative example to demonstrate the major functions.

\subsubsection{Configuring an IMAP connection}
\label{configure-imap}

%Optional: you may include one explanatory video that will appear next to your article, in the right hand side panel. (Please upload any video as a single supplementary file with your article. Only one MP4 formatted, with 50MB maximum size, video is possible per article. Recommended video dimensions are 640 x 480 at a maximum of 30 frames/second. Prior to submission please test and validate your .mp4 file at $ http://elsevier-apps.sciverse.com/GadgetVideoPodcastPlayerWeb/verification$. This tool will display your video exactly in the same way as it will appear on ScienceDirect.).

As we demonstrated in \cref{fig:schema1}, the first step for using mRpostman is to configure an IMAP connection. It consists of creating a connection token object of class \texttt{ImapCon} that will retain all the relevant information to issue requests toward the server.

\textbf{\texttt{configure\_imap}} is the function used to configure and create a new IMAP connection. The mandatory arguments are three character strings: \texttt{url}, \texttt{username}, and \texttt{password} for plain authentication; or \texttt{url}, \texttt{username}, and \texttt{xoauth2\_bearer} for OAuth2.0 authentication\footnote{Please refer to the \textit{``IMAP OAuth2.0 authentication in mRpostman''} vignette in \cite{mRpostman}.}.

The following example illustrates how to configure a connection to a Microsoft Exchange IMAP 4 server; more specifically, to an Office 365 Outlook account using plain authentication.

\begin{verbatim}
library("mRpostman")
con <- configure_imap(url = "imaps://outlook.office365.com",
                      username = "user@agency.gov",
                      password = rstudioapi::askForPassword())
\end{verbatim}

We opted for using an Outlook Office 365 account as an example in order to highlight the difference between mRpostman and the other two CRAN packages which, although also capable of receiving e-mails, are restricted to Gmail accounts and fewer IMAP functionalities. Although mRpostman is able to theoretically connect to any mail provider\footnote{Besides Outlook Office 365, the package has been already successfully tested with Gmail, Yahoo, Yandex, AOL, and Hotmail accounts.}, the Outlook Office 365 service is broadly used by universities and companies. This enriches the range of data analyses applications of this package, thus justifying our choice.

In a hypothetical situation where the user needs to simultaneously connect to more than one e-mail account (in different providers or not) in the same R session, it can be easily attained by creating and configuring multiple connection tokens, such as \texttt{con1}, \texttt{con2}, and so on. 

\subsubsection{Selecting a mail folder}
\label{select-folder}

Mailboxes are structured as folders in the IMAP protocol. This allows us to replicate many of the operations done in a local folder such as creating, renaming or deleting folders. As messages are kept inside the mail folders, users need to select one of them whenever they intend to execute a search, fetch or other message-related operation, as presented in \cref{fig:schema1}.

In this sense, the \texttt{select\_folder} method is one of the key features of this package. It selects a mail folder for the current IMAP section. The mandatory argument is a character string containing the \texttt{name} of the folder to be selected. 

Supposing that we want to select the \texttt{"INBOX"} folder and considering that we are going to use the same connection object (\texttt{con}) that has been previously created, the command would be:

\begin{verbatim}
con$select_folder(name = "INBOX")
\end{verbatim}

Further details on other important mailbox management features are provided in \cite{mRpostman}.

\subsubsection{Message search}
\label{search}

The IMAP protocol is designed to allow the execution of single or multi-criteria queries on the mailboxes. This package implements a vast range of IMAP search commands, which consist of a critical feature for performing data analysis on email content. 

As of its version \texttt{1.0.0}, mRpostman has five types of single-criterion search methods implemented: by date; string; flag, size; and span of time (\texttt{WITHIN} extension)\footnote{The \texttt{WITHIN} extension is not supported by all IMAP servers. A call to the \texttt{list\_server\_capabilities} method will present all the IMAP extensions supported by the mail provider\citep{mRpostman}.}. The custom-search, on the other hand, enables the execution of multi-criteria queries by allowing the combination of two or more types of search. However, in this article, we will focus on the single-criterion search-by-string type.

The \textbf{\texttt{search\_string}} method searches messages that contain a specific string or expression. One or more specific sections of a message, such as the \texttt{TEXT} section or the \texttt{TO} header field, for example, must be specified.

In the following code snippet, we search for messages from senders whose mail domain is \texttt{"@ksu.edu"}.

\begin{verbatim}
ids <- con$search_string(expr = "@ksu.edu", where = "FROM")
\end{verbatim}

The resulting object is a vector containing the matched unique ids (UID) or the message sequence numbers\footnote{More details on the message identification methodology deployed by the IMAP protocol are provided in \cite[]{rfc5322, rfc3501, mRpostman}.} such as presented below:

\begin{verbnobox}[\fontsize{9.5pt}{9.5pt}\selectfont]
[1]  60 145 147 159 332 333 336 338 341 428
\end{verbnobox}

Further details on the other single-search methods and the custom-search method available in this package are provided in \cite{mRpostman}.

\subsubsection{Message fetch}
\label{fetch}

After executing a search query, users may be interested in fetching the full content or some part of the messages indicated in the search results. In this regard, mRpostman implements six types of fetch features: 

\textbf{\texttt{fetch\_body}} Fetches the message body (message’s full content), or an specified MIME level, which can refer to the text or the attachments if there are any.

\textbf{\texttt{fetch\_header}} Fetches the message header, which comprises all the components of the \texttt{HEADER} section of a message. Besides the traditional ones (from, to, cc, subject), it may include several more fields.

\textbf{\texttt{fetch\_metadata}} Fetches the message metadata, which consists of some message's attributes such as the internal date, and the envelope (from, to, cc, and subject fields).

\textbf{\texttt{fetch\_text}} Fetches the message text section, which can comprise attachment MIME levels if applicable.

Each of these methods can be seamlessly integrated into a previous search operation so that the returned ids are used as input for the fetch method. 

Above all, these methods consist of a powerful source of information for performing data analysis on e-mail content. Here, we mimic the extraction of the \texttt{TEXT} portion of a message. Although there is a \texttt{fetch\_text} method, the recommended approach is to use \texttt{fetch\_body(..., mime\_level = 1L)} because the former may collect attachment parts along with the message text.

\begin{verbatim}
out <- ids %>%
  fetch_body(mime_level = 1L)
\end{verbatim}

Once the messages are fetched, the text can be cleaned and decoded with the \texttt{clean\_msg\_text} helper function. A subsequent call to the \texttt{writeLines} base R function produces a clean printing of the fetched text:

\begin{verbatim}
cleaned_text <- clean_msg_text(msg_list = out)
writeLines(cleaned_text[[1]])
\end{verbatim}

\begin{verbnobox}[\fontsize{9.5pt}{9.5pt}\selectfont]
 Receipt Number: XXXXXXX
 Customer: Vieira de Castro Quadros, Allan
 Kansas State University
 Current Date: 04/15/2020

 Description                                                               Amount
 --------------------------------------------------------------------------------
 HOUSING & DINING                                                          $30.00
      User Number: XXXXXXXXX                                                     
                                                             Total         $30.00

 Payments Received                                                         Amount
 --------------------------------------------------------------------------------
 07 CREDIT CARD PAYMENTS                                                   $30.00
      Visa XXXXXXXXXXXX8437
      Authorization # XXXXXX
                                                             Total         $30.00

 Thank you for the payment.
\end{verbnobox}

Besides other applications, the exported function \texttt{clean\_msg\_text} can be used to decode hexadecimal and base 64 characters in the text and other parts of the message. In some locales such as French, German or Portuguese speaking countries, message parts may contain non-ASCII characters. SMTP servers, then, encode it using the RFC 2047 specifications when sending the e-mail. In these cases, \texttt{clean\_msg\_text} is capable of correctly decoding the non-ASCII characters.

\subsubsection{Attachment extraction}
\label{attach}

In its pretension to be an IMAP client for R, mRpostman provides methods that enable users to list and download message payloads. This feature can be particularly critical for automating the analysis of attachment data files, for instance.

Attachments can be downloaded using two different approaches in this package: extending the \texttt{fetch\_text/body} operation by adding an attachment extraction step at the end of the workflow with \texttt{get\_attachments}; or directly fetching attachment parts via the \texttt{fetch\_attachments} method. In this article, we focus on the first type of attachment methods, adding a step to our previous workflow. 

The \texttt{get\_attachments} method extracts attachment files from the fetched messages and saves these files to the disk. In the following code excerpt, we extract attachments in a unique pipeline that gathers fetching and search steps.

\begin{verbatim}
con$search_string(expr = "@ksu.edu", where = "FROM") %>%
  con$fetch_text() %>%
  con$get_attachments()
\end{verbatim}

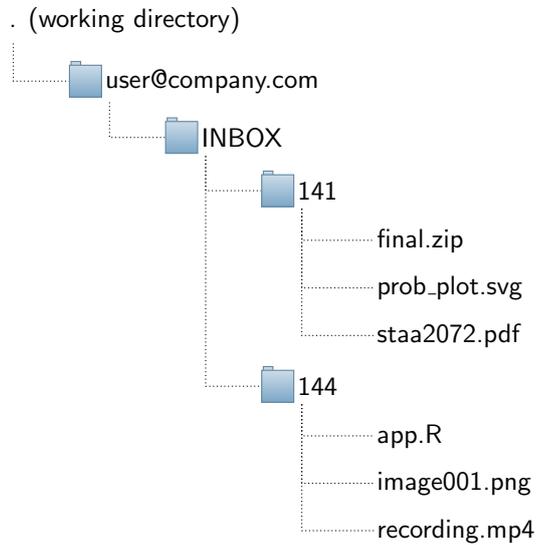
\begin{figure} 
  \centering
  \footnotesize
    \begin{forest}
        for tree={font=\sffamily, grow'=0,
        folder indent=.9em, folder icons,
        edge=densely dotted}
    [. (working directory)
    [user@company.com
    [INBOX
        [141
            [final.zip, is file]
            [prob\_plot.svg, is file]
            [staa2072.pdf, is file]
        ]
        [144
            [app.R, is file]
            [image001.png, is file]
            [recording.mp4, is file]
        ]
    ]
    ]
    ]
  %}
    \end{forest}
  \caption{Local directory tree for the extracted attachment files}
  \label{fig:tree}
\end{figure}

During the execution, the software locally saves the extracted attachments into sub-folders inside the user's working directory. These sub-folders are named following the messages' ids. The attachments are placed into their respective messages' sub-folders as demonstrated in \cref{fig:tree}. Note that the parent levels are named after the informed username and the selected mail folder.

For more information on the other attachment-related methods, the reader should refer to the documentation in \cite{mRpostman}.

\section{Illustrative Examples}
\label{examples}

To demonstrate the capabilities of the proposed software, we explore two use cases of this package in support of data analysis tasks: a simple study of the frequency of e-mails grouped by senders; and a sentiment analysis run on a set of e-mails received during a period. The R scripts needed for reproducing these examples are provided in the appendixes. Although the results cannot be exactly reproduced once it reflects the author's mailbox contents, they can be easily adapted to the reader's context.

\subsection{Frequency analysis of e-mail data}
\label{example1}

\begin{figure}[H]
  \includegraphics[width=1.0\linewidth]{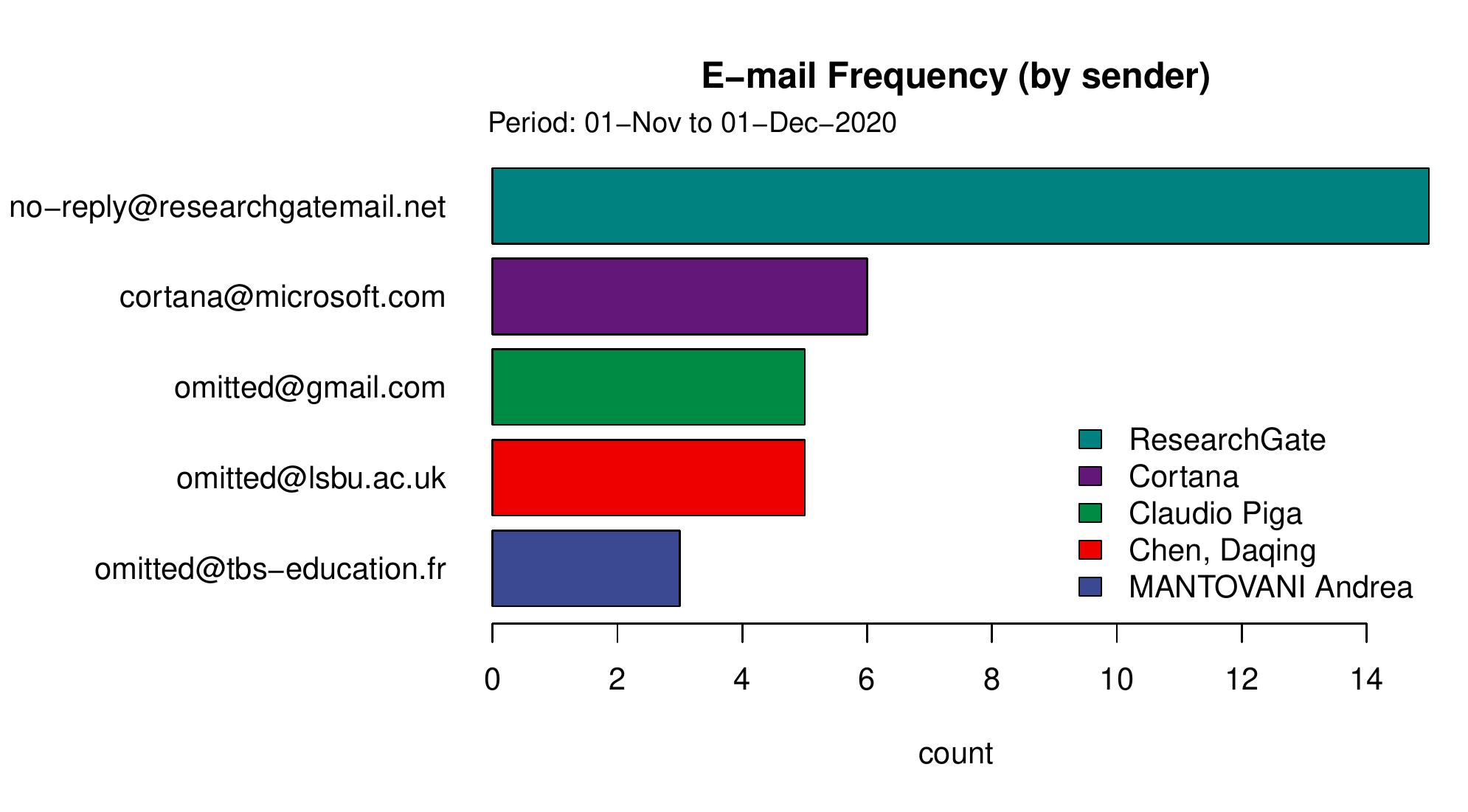}
  \caption{An example of e-mail frequency analysis grouped by sender}
  \label{fig:frequency-ex1}
\end{figure}

In the first example, we run a simple analysis of the e-mail frequency with regard to senders. This can be especially useful in professional fields, such as marketing and customer service offices. A period of analysis was defined, and a search-by-date is performed using the \texttt{search\_period} method. Then, senders' information for the returned ids are fetched via \texttt{fetch\_metadata}, using the \texttt{ENVELOPE} attribute. After some basic manipulation with regular expressions, the data is ready to be plotted as shown in \cref{fig:frequency-ex1}.

The same kind of analysis can be replicated for the messages' subjects with only a few modifications in the regular expressions code chunks. Considering that some companies/users deal with subject-standardized e-mails, this approach can be useful to analyze the frequency of e-mails with regard to different categories of subjects.

\subsection{Sentiment analysis on e-mail data}
\label{example2}

For the sentiment analysis example, we also define a period of analysis and run a \texttt{search\_period} query. Then, we retrieve the text part of the messages by fetching the first MIME level with \texttt{fetch\_body(..., mime\_level = 1L)}. The texts go trough a first cleaning step with a call to the \texttt{clean\_msg\_text} function. After further cleaning procedures, we use a lexicon\citep{saif10} via the syuzhet package\citep{syuzhet} to evaluate the sentiment of each e-mail. The output below is a subset of the resulting data frame. The last two columns indicate, respectively, the counts of negative and positive words for each message. The other columns provide counts related to detailed emotions, which are not necessarily positive nor negative.

\begin{verbnobox}[\fontsize{9.5pt}{9.5pt}\selectfont]
        anger anticipation disgust fear joy sadness surprise trust negative positive
body91      1            5       1    1   2       2        0     9        1       13
body92      0            1       0    0   1       0        0     3        0        1
body93      0            3       0    2   0       1        2     2        1        3
body94      0            1       0    1   0       0        1     4        1        4
body95      0            5       0    0   3       0        2     8        0       13
body96      0            0       0    0   0       0        0     0        0        0
body97      4           20       4   11  13      11        4    25       16       51
body98      0            3       0    0   2       0        1     4        0        6
body99      3            9       1    6   1       5        2    16       14       24
body100     4           15       1   13   6       7        6    15       16       31
\end{verbnobox}

\section{Impact}
\label{impact}

As we have demonstrated, mRpostman clearly fills an existent gap of a broad, complete, and, at the same time, easy-to-use IMAP client for the R language. The package has consolidated itself as an important tool for collecting massive e-mail content, thus contributing to data analysis tasks in R.

Although all sort of users have been taking advantage of this package, we are inclined to think that its use has been prevailing amid companies. We have received a considerable number of feedback from enterprise users who deploy mRpostman as an additional feature for automatically producing daily reports based on attachment data files. Besides this, there are important applications for marketing and post-sales departments, for example. They can also deploy this package to collect e-mail data for analyzing e-mail frequency, or performing sentiment analysis, as we have demonstrated in Section \ref{examples}.

%\textbf{This is the main section of the article and the reviewers weight the description here appropriately}

%Indicate in what way new research questions can be pursued as a result of the software (if any).

%Indicate in what way, and to what extent, the pursuit of existing research questions is improved (if so).

%Indicate in what way the software has changed the daily practice of its users (if so).

%Indicate how widespread the use of the software is within and outside the intended user group.

%Indicate in what way the software is used in commercial settings and/or how it led to the creation of spin-off companies (if so).

\section{Conclusions}
\label{conclusions}

mRpostman aims to provide an easy-to-use IMAP client for R. Its design allows the efficient, elegant, and intuitive execution of several IMAP commands on a wide range of mail providers. Consequently, users cannot only manage their mailboxes but also conduct e-mail data analysis from inside R. Finally, because IMAP is such a complex protocol, this package is in constant development, which means that new features are to be implemented in future versions.

\section{Conflict of Interest}

No conflict of interest exists:
We wish to confirm that there are no known conflicts of interest associated with this publication and there has been no significant financial support for this work that could have influenced its outcome.

\section*{Acknowledgements}
\label{acknowledgements}

The author would like to acknowledge the Department of Statistics at Kansas State University (K-State) for the assistantship provided for his doctorate studies. He wants to especially thank Dr. Christopher Vahl and Dr. Michael Higgins for the academic support. The author also acknowledges the academic guidance of Dr. George von Borries at the University of Brasilia (UnB). The contents of this article are the responsibility of the author and do not reflect the views of K-State or UnB.

\appendix

\section{Code for example 1}
\label{appendixA}

\begin{verbnobox}[\fontsize{9.5pt}{9.5pt}\selectfont]
    
library(mRpostman)

con <- configure_imap(
  url="imaps://outlook.office365.com",
  username="user@company.com",
  password=rstudioapi::askForPassword()
)

con$select_folder(name = "INBOX")

meta_res <- con$search_period(since_date_char = "01-Nov-2020",
                              before_date_char = "01-Dec-2020") %>%
  con$fetch_metadata(attribute = "ENVELOPE")

# cleaning
# step 1
clean_meta <- lapply(meta_res, function(x){
  regmatches(x, regexpr(pattern = "\\(\\(.*\"(.*?)\"\\)\\)", x, perl = TRUE))
})

# step 2
# cleaning Ccs
senders1 <- lapply(clean_meta, function(x){
  gsub(")) NIL .*$|)) .*$|))$", "", x)
})

# step 3
senders1 <- lapply(senders1, function(x){
  gsub('^\\(\\(|\"+', "", x)
})

# splitting
name <- c()
email <- c()

for (i in seq_along(senders1)){
  # i = 1
  out <- unlist(strsplit(senders1[[i]], " NIL "))
  name <- c(name, out[1])
  email <- c(email, gsub(" ", "@", out[2]))
}

df <- data.frame(name, email)
df$name <- decode_mime_header(string = as.character(df$name))
df2 <- as.data.frame(table(df$email))
colnames(df2) <- c("email", "count")

df2 <- df2[order(-df2[,2]), ][1:5,]
df2$name <- unique(df$name[df$email %in% df2$email])

par(mar=c(5,13,4,1)+.1)
pal_cols <- c('#3B4992FF', '#EE0000FF', '#008B45FF', '#631879FF', '#008280FF')
barplot(rev(df2$count),
        main = "E-mail Frequency (by sender)",
        xlab = "count",
        names.arg = rev(df2$email),
        las = 1,
        col = pal_cols,
        horiz = TRUE)
mysubtitle <- "Period: 01-Nov to 01-Dec-2020"
legend(x = "bottomright", legend = df2$name, fill = rev(pal_cols), bty = "n", y.intersp = 1)
mtext(side=3, line=0.3, at=-0.07, adj=0, cex=0.9, mysubtitle)

\end{verbnobox}

\section{Code for example2}
\label{appendixB}

\begin{verbnobox}[\fontsize{9.5pt}{9.5pt}\selectfont]

library(mRpostman)
con <- configure_imap(url="imaps://outlook.office365.com",
                      username="user@company.com",
                      password=rstudioapi::askForPassword(),
                      timeout_ms = 20000
)

con$select_folder("INBOX")

ids <- con$search_period(since_date_char = "10-Oct-2020", before_date_char = "20-Dec-2020")

fetch_res2 <- ids %>%
  con$fetch_body(mime_level = 1L)

cleaned_text_list <- clean_msg_text(msg_list = fetch_res2)

cleaned_text_list[[4]]

for (i in seq_along(cleaned_text_list)) {
  
  clean_text <- gsub("\r\n", " ", cleaned_text_list[[i]])
  
  clean_text <- unlist(strsplit(clean_text, " "))
  
  words <- clean_text[!grepl("\\d|_|http|www|nbsp|@|(?<=[[:lower:]])(?=[[:upper:]])",
                             clean_text, perl = TRUE)]
  
  words <- tolower(gsub("\\W+", "", words))
  
  words <- gsub('[^a-zA-Z|[:blank:]]', "", words)
  
  cleaned_text_list[[i]] <- paste(words, collapse = " ")

}

cleaned_text_df <- do.call('rbind', cleaned_text_list)
library(syuzhet)
email_sentiment_df <-get_nrc_sentiment(cleaned_text_df)
rownames(email_sentiment_df) <- rownames(cleaned_text_df)
head(email_sentiment_df,10)

\end{verbnobox}



%% References:


%\begin{thebibliography}{00}


%\bibitem{}

%\end{thebibliography}

\bibliography{references}

\begin{thebibliography}{10}
\expandafter\ifx\csname url\endcsname\relax
  \def\url#1{\texttt{#1}}\fi
\expandafter\ifx\csname urlprefix\endcsname\relax\def\urlprefix{URL }\fi
\expandafter\ifx\csname href\endcsname\relax
  \def\href#1#2{#2} \def\path#1{#1}\fi

\bibitem{R}
{R Core Team}, \href{https://www.R-project.org/}{R: A Language and Environment
  for Statistical Computing}, R Foundation for Statistical Computing, Vienna,
  Austria (2020).
\newline\urlprefix\url{https://www.R-project.org/}

\bibitem{rmarkdown}
J.~Allaire, Y.~Xie, J.~McPherson, J.~Luraschi, K.~Ushey, A.~Atkins, H.~Wickham,
  J.~Cheng, W.~Chang, R.~Iannone,
  \href{https://github.com/rstudio/rmarkdown}{rmarkdown: Dynamic Documents for
  R}, r package version 2.5 (2020).
\newline\urlprefix\url{https://github.com/rstudio/rmarkdown}

\bibitem{BOI08}
P.~Heinlein, P.~Hartleben, The Book of IMAP: Building a Mail Server with
  Courier and Cyrus, No Starch Press, 2008.

\bibitem{sendmailR}
O.~Mersmann, \href{https://CRAN.R-project.org/package=sendmailR}{sendmailR:
  send email using R}, r package version 1.2-1 (2014).
\newline\urlprefix\url{https://CRAN.R-project.org/package=sendmailR}

\bibitem{mailR}
R.~Premraj, \href{https://CRAN.R-project.org/package=mailR}{mailR: A Utility to
  Send Emails from R}, r package version 0.4.1 (2015).
\newline\urlprefix\url{https://CRAN.R-project.org/package=mailR}

\bibitem{mail}
L.~Himmelmann, \href{https://CRAN.R-project.org/package=mail}{mail: Sending
  Email Notifications from R}, r package version 1.0 (2011).
\newline\urlprefix\url{https://CRAN.R-project.org/package=mail}

\bibitem{blatr}
S.~M. Bache, \href{https://CRAN.R-project.org/package=blatr}{blatr: Send Emails
  Using 'Blat' for Windows}, r package version 1.0.1 (2015).
\newline\urlprefix\url{https://CRAN.R-project.org/package=blatr}

\bibitem{gmailr}
J.~Hester, \href{https://CRAN.R-project.org/package=gmailr}{gmailr: Access the
  'Gmail' 'RESTful' API}, r package version 1.0.0 (2019).
\newline\urlprefix\url{https://CRAN.R-project.org/package=gmailr}

\bibitem{blastula}
R.~Iannone, J.~Cheng,
  \href{https://CRAN.R-project.org/package=blastula}{blastula: Easily Send HTML
  Email Messages}, r package version 0.3.2 (2020).
\newline\urlprefix\url{https://CRAN.R-project.org/package=blastula}

\bibitem{emayili}
A.~B. Collier, \href{https://CRAN.R-project.org/package=emayili}{emayili: Send
  Email Messages}, r package version 0.4.4 (2020).
\newline\urlprefix\url{https://CRAN.R-project.org/package=emayili}

\bibitem{edeR}
A.~B. Collier, \href{https://CRAN.R-project.org/package=edeR}{edeR: Email Data
  Extraction Using R}, r package version 1.0.0 (2014).
\newline\urlprefix\url{https://CRAN.R-project.org/package=edeR}

\bibitem{rfc3501}
M.~Crispin, \href{https://tools.ietf.org/html/rfc3501}{Internet message access
  protocol - version 4rev1}, request for Comments 3501 (RFC 3501), Internet
  Engineering Task Force (IETF) (2003).
\newline\urlprefix\url{https://tools.ietf.org/html/rfc3501}

\bibitem{rfc2047}
K.~Moore, \href{https://tools.ietf.org/html/rfc2047}{{Multipurpose Internet
  Mail Extensions (MIME)}, part three: Message header extensions for non-ascii
  text}, request for Comments 2047 (RFC 2047), Internet Engineering Task Force
  (IETF) (1996).
\newline\urlprefix\url{https://tools.ietf.org/html/rfc2047}

\bibitem{magrittr}
S.~M. Bache, H.~Wickham,
  \href{https://CRAN.R-project.org/package=magrittr}{magrittr: A Forward-Pipe
  Operator for R}, r package version 1.5 (2014).
\newline\urlprefix\url{https://CRAN.R-project.org/package=magrittr}

\bibitem{R6}
W.~Chang, \href{https://CRAN.R-project.org/package=R6}{R6: Encapsulated Classes
  with Reference Semantics}, r package version 2.5.0 (2020).
\newline\urlprefix\url{https://CRAN.R-project.org/package=R6}

\bibitem{curl20}
J.~Ooms, \href{https://CRAN.R-project.org/package=curl}{curl: A Modern and
  Flexible Web Client for R}, r package version 4.3 (2020).
\newline\urlprefix\url{https://CRAN.R-project.org/package=curl}

\bibitem{libcurl20}
D.~Stenberg, \href{https://curl.haxx.se/}{libcurl - the multiprotocol file
  transfer library}, version 7.69.1 (2020).
\newline\urlprefix\url{https://curl.haxx.se/}

\bibitem{mRpostman}
A.~Quadros, \href{https://allanvc.github.io/}{mRpostman: An IMAP Client for R},
  r package version 1.0.0 (2020).
\newline\urlprefix\url{https://allanvc.github.io/}

\bibitem{rfc5322}
P.~Resnick, \href{https://tools.ietf.org/html/rfc5322}{Internet message
  format}, request for Comments 5322 (RFC 5322), Internet Engineering Task
  Force (IETF) (2008).
\newline\urlprefix\url{https://tools.ietf.org/html/rfc5322}

\bibitem{saif10}
S.~Mohammad, P.~Turney,
  \href{http://saifmohammad.com/WebPages/lexicons.html}{Emotions evoked by
  common words and phrases: Using mechanical turk to create an emotion
  lexicon}, in: CAAGET '10: Proceedings of the NAACL HLT 2010 Workshop on
  Computational Approaches to Analysis and Generation of Emotion in Text, Los
  Angeles, California, 2010, p. 26–34, june, 2010.
\newline\urlprefix\url{http://saifmohammad.com/WebPages/lexicons.html}

\bibitem{syuzhet}
M.~L. Jockers, \href{https://github.com/mjockers/syuzhet}{Syuzhet: Extract
  Sentiment and Plot Arcs from Text}, r package version 1.0.4 (2015).
\newline\urlprefix\url{https://github.com/mjockers/syuzhet}

\end{thebibliography}

%Please add the reference to the software repository if DOI for software  is available. 

%Ancillary data table required for sub version of the executable software: (x.1, x.2 etc.) kindly replace examples in right column with the correct information about your executables, and leave the left column as it is.

%\begin{table}[!ht]
%\resizebox{\textwidth}{!}{%
%\begin{tabular}{|l|p{6.5cm}|p{6.5cm}|}
%\hline
%\textbf{Nr.} & \textbf{(Executable) software metadata description} & \textbf{Please fill in this column} \\
%\hline
%S1 & Current software version & 1.0.0 \\
%\hline
%S2 & Permanent link to executables of this version  & https://cloud.r-project.org/web/packages/mRpostman/index.html \\
%\hline
%S3 & Legal Software License & GPL-3 \\
%\hline
%S4 & Computing platforms/Operating Systems & Linux, OS X, Microsoft Windows, Solaris. \\
%\hline
%S5 & Installation requirements \& dependencies & R environment version 3.5.3 and up; libcurl: libcurl-devel (rpm) or libcurl4-openssl-dev (deb); R packages: curl, R6, stringr, stringi, magrittr, assertthat, base64enc, utils, rvest, xml2\\
%\hline
%S6 & If available, link to user manual - if formally published include a reference to the publication in the reference list & https://cloud.r-project.org/web/packages/mRpostman/mRpostman.pdf \\
%\hline
%S7 & Support email for questions & quadros@ksu.edu\\
%\hline
%\end{tabular}}
%\caption{Software metadata (optional)}
%\label{metadata2} 
%\end{table}

\end{document}